\documentstyle[12pt,epsfig]{article}

\def\epr{\varepsilon^\prime/\varepsilon}
\def\repr{\mathrm{Re}\,(\epr)}
\def\ko{K^0}
\def\kobar{\overline{K^0}}
\def\kl{K_L}
\def\ks{K_S}
\def\pipin{\pi^0\pi^0}
\def\pipic{\pi^+\pi^-}
\def\kethree{K_{e3}}
\def\kmuthree{K_{\mu3}}

\def\Title#1{\begin{center} {\Large {\bf #1} } \end{center}}

\begin{document}

\begin{center} 
\textsc{Proceedings of the 1999 Lepton-Photon Symposium, Stanford, U.S.A.} 
\end{center}
\medskip

\Title{Results on direct CP-violation from NA48}

\bigskip\bigskip


\begin{raggedright}  

{\it Giles Barr\index{Barr, G.D.}\\
NA48 Collaboration\footnotemark[1]\\
CERN, Geneva, Switzerland } 
\bigskip\bigskip
\end{raggedright}

\section{Introduction}

A long-standing question in high energy physics has been the origin of
the phenomenon of CP violation\index{CP violation}.  CP violation was
first observed in the decay $\kl \rightarrow \pipic$~\cite{cronin}.
Effects have subsequently been found in $\kl \rightarrow
\pipin$~\cite{cppipin}, the charge asymmetry of $\kl \rightarrow e^\pm
\pi^\mp\nu$ ($\kethree$)~\cite{cpkethree} and $\kl \rightarrow \mu^\pm
\pi^\mp\nu$ ($\kmuthree$)~\cite{cpkmuthree}, $\kl \rightarrow
\pipic\gamma$~\cite{cppipicgamma} and most recently in $\kl
\rightarrow \pi\pi ee$~\cite{cppipiee}.  All of these effects can be
explained by applying a single CP violating effect in the mixing
between $\ko$ and $\kobar$ which proceeds through the box Feynman
diagram and are characterised by the parameter $\varepsilon$.

A second form of CP-violation with different characteristics can also
be investigated in the decay of the neutral K-mesons.  The $CP = -1$
kaon state (the $K_2$) can decay directly into a $2\pi$ final state
without first mixing into a kaon with $CP = +1$.  This is referred to
as direct CP-violation, may proceed by the penguin Feynman diagram and
is characterised by the parameter $\varepsilon^\prime$.  The measured
quantity is the double ratio of the decay widths, which, in the NA48
experiment is equivalent to the double ratio of four event-counts
\begin{eqnarray}
 R & \equiv &
   \frac{\Gamma(\kl \rightarrow \pipin) /\, \Gamma(\ks \rightarrow \pipin)}
        {\Gamma(\kl \rightarrow \pipic) /\, \Gamma(\ks \rightarrow \pipic)}
     \nonumber \\
   & = & 
   \frac{N(\kl \rightarrow \pipin) /\, N(\ks \rightarrow \pipin)}
        {N(\kl \rightarrow \pipic) /\, N(\ks \rightarrow \pipic)}
     \label{eqnr} \\
    & \simeq & 1-6\times \repr. \nonumber
\end{eqnarray}
A non-zero value of $\repr$ signals that direct CP violation exists. 

CP-violating effects are also being searched for in the decays of
B-mesons~\cite{lp99bsession}.

This talk describes the first results on $\repr$ from the NA48\index
{NA48 experiment} experiment at CERN.  More precise details of this
measurement, in particular on the data selection and corrections have
since been published in ref.~\cite{pleprime}.

The previous round of measurements on direct CP violation ended with
some disagreement.  The NA31 experiment at CERN published the first
evidence for direct CP violation~\cite{hburkhardt} and their final
measured value was $(23.0 \pm 6.5) \times 10^{-4}$~\cite{na31}, while
the E731 experiment at Fermilab measured $(7.4 \pm 5.9) \times
10^{-4}$~\cite{e731} which is consistent with no effect.  A new
measurement of $(28.0 \pm 4.1) \times 10^{-4}$ from the KTeV
collaboration~\cite{ktev} is the subject of another talk at this
conference~\cite{lp99ktev}.

\section{Principle of the experiment}

The measurement of $\repr$ proceeds by counting the number of decays
from a $\ks$ beam into both $\pipic$ and $\pipin$ within a given
energy range and fiducial region and counting the number of decays
into the same modes from a $\kl$ beam in the same energy and fiducial
region.  These four numbers are then corrected for background,
acceptance and the influence of uncorrelated activity in the detectors
and are used to calculate $R$ as defined in equation (\ref{eqnr}).  By
collecting the data for both the $\pipic$ and $\pipin$ events
simultaneously, the variation of the flux of particles cancels when
inserting the numbers into (\ref{eqnr}).

To simplify the task of correcting for the influence of uncorrelated
activity in the detectors, the data are collected with both the $\ks$
and $\kl$ beams switched on simultaneously.  Most of the uncorrelated
activity in the detectors originates from the $\kl$ beam, and so the
rate of uncorrelated activity is, to a good approximation, the same at
the time when either a $\ks$ or a $\kl$ decay occurs.  This principle
applies both for losses where the uncorrelated activity causes the
reconstruction of the event to fail and also for cases where pile-up
of signals in the detector cause trigger inefficiencies.

Similarly, the acceptance correction is simplified by arranging for
the $\ks$ and $\kl$ events to both be collected with the same decay
profile along the beam direction.  By simply counting the events, the
$\ks$ decays have a distribution characterised by an exponential decay
corresponding to the $\ks$ lifetime of $c\tau = 2.68$~cm ($\gamma
c\tau \sim 5$~m), while the $\kl$ distribution is almost completely
flat.  Instead, in the NA48 analysis, a weight is applied to the $\kl$
events as a function of the measured decay position along the beam in
order to give them the same exponential decay distribution as the
$\ks$.  Consequently, the geometrical acceptance is very similar for
$\ks \rightarrow \pipic$ and $\kl \rightarrow \pipic$ and the
correction to the ratio cancels in the ratio in (\ref{eqnr}).  The
same principle is also applied to the $\pipin$ decays.

The kaon beams are produced from a proton beam which strikes two
targets, producing particles of different types and energies.  The
charged particles are swept away with a magnet and dumped leaving only
a neutral beam containing $\ks$, $\kl$, $\Lambda$, neutrons, photons
and neutrinos which passes through the collimators.  The kaons which
are used for the measurement have energies between 70 and 170~GeV.
The $\kl$ target is placed 120~m upstream of the fiducial region where
the events are counted in order to let $\ks$ and $\Lambda$ decay.  The
$\ks$ target is placed as close as possible to the fiducial region,
the gap which is needed for the sweeping magnet and charged particle
dump is made as small as possible.  This gap changes the $\ks$ energy
spectrum due to Lorrenz contraction. This change is the same for
$\pipic$ and $\pipin$ decays and so cancels in principle in
(\ref{eqnr}).  However, since the acceptance as a function of the kaon
energy is different in the two modes, the analysis is done in 5~GeV
wide bins of energy such that the product of the acceptance and the
energy spectrum is constant across each energy bin and the
cancellation of the energy spectra differences occurs in (\ref{eqnr}).

\section{The experimental apparatus}

As described above, the experiment consists of two beams of kaons, one
from a target 120~m upstream for measuring $\kl$ decays (which must be
of high intensity because $\kl$ CP violating decays are rare) and one
from a target immediately in front of the fiducial region for measuring
the $\ks$ decays (low intensity, since the $\ks$ decays predominantly
to the two-body final states).  This is followed by a detector system
for measuring $\pipic$ and $\pipin$ final states and separating them
from the large rate of background from three-body decays from the
$\kl$ beam.  In addition, a `tagging' system is required in order to
determine from which beam each decay occurred.

The kaon beams are both produced from 450~GeV protons from the slow
extraction of the CERN super proton synchrotron (SPS), a burst lasting
2.4~s is produced every 14.4~s.  The 200~MHz time structure from the
accelerator is almost completely removed by debunching before the
protons are extracted from the accelerator.  The proton intensities on
the $\kl$ and $\ks$ targets are $1.1 \times 10^{12}$ and $3 \times
10^7$ protons per burst respectively.  The protons for the $\ks$ beam
are derived from the protons which passed through the $\kl$ target
without interacting.  A small fraction of these are deflected by channelling
using a bent crystal (to separate them from the debris from the
target) and transported to the $\ks$ target.  At the entrance to the
fiducial region, the two beams are separated by 6.8~cm.  The $\ks$
beam is tilted slightly so that the two beams converge with an angle
of 0.6~mrad at the detector (This makes the detector
illumination as similar as possible in the two modes).

\begin{figure}[htb]
  \begin{center}
    \resizebox{0.9\textwidth}{!}{
      \rotatebox{0}{
        \includegraphics[width=\linewidth]{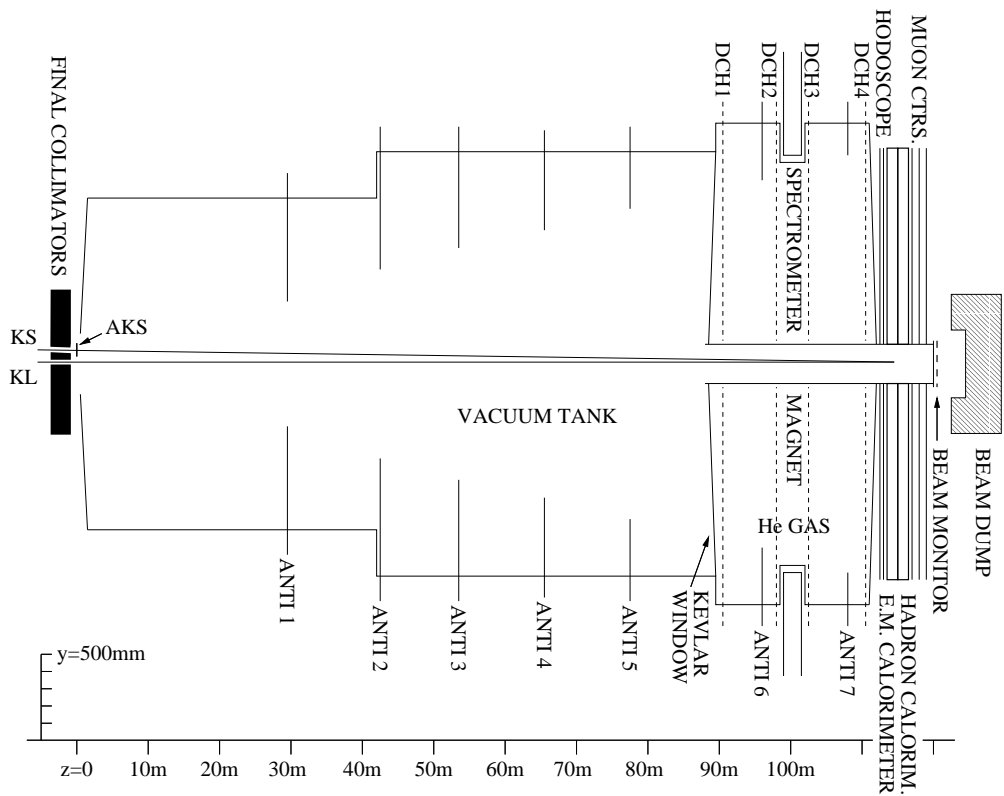}
        }
      }
    \caption{Schematic layout of the NA48 detector.}
    \label{figlayout}
  \end{center}
\end{figure}

The layout of the detector is shown in fig.~\ref{figlayout}.  The
$\pipin$ events are measured in a liquid krypton electromagnetic
(e.m.)  calorimeter.  The 13212 readout cells are in a tower
arrangement so that the distribution of energy is measured by flash
ADCs in a Cartesian grid of 2~cm $\times$ 2~cm cells $\times$ 25~ns.
The calorimeter is calibrated with an electronic pulser and with
electrons from $\kethree$ events.  The events are triggered by a
second system of flash ADCs which measure the signals on summed
readout cells whose output is fed into a digital computation pipeline.
A trigger decision is generated every 25~ns based on a minimum total
energy, a maximum proper decay time and the number of peaks.  The
$\pipic$ events are detected in a charged particle spectrometer
consisting of four multiwire drift chambers and a magnet with a
horizontal momentum kick of 265~MeV/$c$.  The e.m.~calorimeter and a
set of three muon detectors are used for particle identification to
distinguish $\kethree$ and $\kmuthree$ events respectively from the
signal events.  The triggering of the $\pipic$ events occurs in
several steps: the first level is based on hits in a pair of planes of
scintillation counter hodoscopes in coincidence with an energy
deposition in the calorimeters (a hadron calorimeter is located
downstream of the LKr for this purpose); the second level is based on
a reconstruction of the hits in the spectrometer to determine the kaon
energy, kaon mass and proper decay time in an array of
microprocessors.  The beam pipe runs down the middle of all the
detectors to allow the neutrons, photons and undecayed kaons to pass
through the detector to a beam dump downstream.

The tagging to assign each measured decay to either the $\ks$ or $\kl$
beams is performed by determining whether or not a proton hit the
$\ks$ target in coincidence with the decay.  The event time of
$\pipic$ events is reconstructed from the hits in the hodoscope, the
event time of $\pipin$ events is reconstructed from the digitised
pulse profiles for the LKr cells near the centre of each shower.  A
series of scintillation counters (the tagger) is installed along the
path of the protons for the $\ks$ beam.  If a coincidence ($\pm 2$~ns
window) between the event time and a proton reconstructed in the
tagger occurs, the event is assigned to be a $\ks$, otherwise it is
assigned to be a $\kl$.  For $\pipic$ events, the vertex resolution 
from the drift chambers is good enough to directly see from which
target the kaon was produced (vertex tagging).

\section{Analysis}

The analysis is performed in twenty 5~GeV wide bins of kaon energy and
the corrections are applied to each bin individually (except where
indicated).  The analysis proceeds by counting the number of candidate
events in each of the four decay channels and applying the weights
described above to the $\kl$ events to synthesise the exponential
decay distribution of the $\ks$ events ($\tau$-weighting).

{\bf Acceptance:} Corrections are made for the small residual
acceptance difference after the weighting which comes principally from
the 0.6~mrad angular separation between the beams and the different
beam divergences (0.15~mrad in $\kl$ and 0.375~mrad in $\ks$) leading
to a total acceptance correction to $R$ of $(+29 \pm 12) \times
10^{-4}$.

{\bf Energy scale and non-linearity:} The decay position $z$ of the
events must be reconstructed accurately on an absolute scale so that
the definition of the fiducial region is precisely the same for
charged and neutral events.  The length of the fiducial region is
defined in terms of the proper decay time and a cut is made at
3.5~$\tau_S$ where $\tau_S$ is the mean $\ks$ lifetime from the
PDG~\cite{pdg}.  This corresponds to 13.2 and 32.0~m at 70 and 170~Gev
respectively.  The decay position of $\pipic$ events is determined
from geometry. The decay position of $\pipin$ events is
determined from the energies and relative positions of the
electromagnetic clusters and a shift in the energy scale will cause a
corresponding shift in the decay vertex position and lead to an
uncertainty on $R$.  A counter in the $\ks$ beamline (shown as AKS on
figure~\ref{figlayout}) is used to veto kaon decays which decay
upstream.  This is used to define the start of the fiducial region in
the $\ks$ mode and also provides a fiducial mark in the decay
distribution (it marks the upstream end of the fiducial region) which
is used to tune the energy scale factor.  A single factor is used, common
to all energy bins.  This now sets the scale which is used to
define the fiducial region in the $\kl$ beam, the position from which
the $\kl$ $\tau$-weighting starts and the end of the fiducial region
in the $\ks$ mode.

Overall, the sensitivity of the result to the energy scale is low
because of a cancellation in $\kl$ between a stretch of the
fiducial region and the determination of the $\tau$-weights from the
reconstructed vertex position.  Non-linearities in the response of the
electromagnetic calorimeter also play a role in limiting the precision
of the value of $R$.  The e.m.-calorimeter's response as a function of
energy is studied carefully with a variety of methods: electrons from
$\kethree$ decays, special runs in which a $\pi^-$ beam is introduced
into the experiment and strikes a target at a known fixed location
producing photons from $\pi^0 \rightarrow \gamma\gamma$, $\eta
\rightarrow \gamma\gamma$ and $\eta \rightarrow 3\pi^0 \rightarrow
6\gamma$ and using the reconstructed $\pi^0$ masses in $K \rightarrow
2\pi^0$ decays.  The uncertainty in the value of $R$ from energy scale
and non-linearity effects is $\pm 12 \times 10^{-4}$.

{\bf Uncorrelated activity:} A sampling of the uncorrelated activity in
the detectors is obtained by triggering at a random instant in time
but proportional to the intensity of the beams (random events), using
counters near the beam dumps.  The response of the detectors to the
combination of a normal event and a random event is computed by adding
the signals channel by channel and feeding the resulting event through
the reconstruction program.  These events have thereby artificially
been given double the amount of uncorrelated activity and are used to
compute the effect on $R$ leading to a correction of $(-2 \pm 14)
\times 10^{-4}$.  (This correction is not applied to energy bins
separately).

{\bf Backgrounds:} The event samples are background subtracted.
Background to $\pipic$ events from $\kl \rightarrow \pipic\pi^0$ and
from $\Lambda$ decays is completely removed by the analysis cuts.  A
subtraction is performed in $\kl \rightarrow \pipic$ from $\kethree$
and $\kmuthree$ events from the $\kl$ beam by studying events with
reconstructed masses adjacent to the kaon mass and events with a
non-zero transverse momentum. This leads to a correction to $R$ of
$(+23 \pm 4) \times 10^{-4}$.  The background in the $\pipin$ mode
comes exclusively from $\kl \rightarrow 3\pi^0$ decays.  A subtraction
is made based on an extrapolation of events in which the four photons
which were detected can be reconstructed to a mass which is close but
not equal to the $2 \times \pi^0$ configuration.  The extrapolation is
done using a Monte-Carlo simulation leading to a correction to $R$ of
$(-8 \pm 2) \times 10^{-4}$.  An additional correction to $R$ of $(-12
\pm 3) \times 10^{-4}$ is made for the loss of $\pipic$ events which
scatter on the $\kl$ beam collimator system and subsequently
regenerate.

\begin{figure}[htb]
  \begin{center}
    \resizebox{0.8\textwidth}{!}{
      \rotatebox{0}{
        \includegraphics[width=\linewidth]{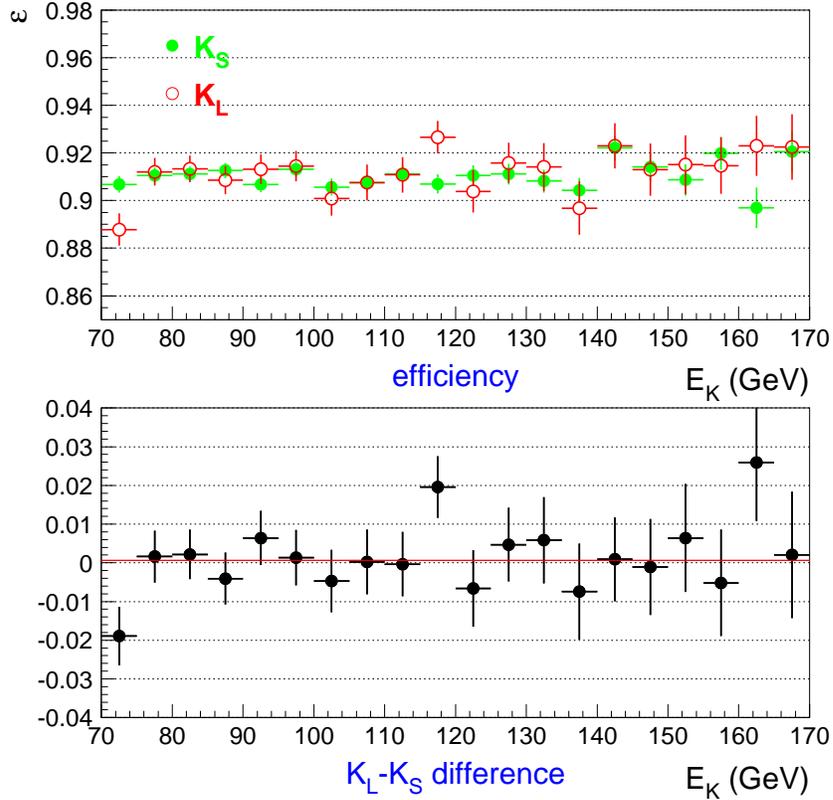}
        }
      }
    \caption{Charged trigger efficiency as a function of kaon energy.}
    \label{figtrig}
  \end{center}
\end{figure}

{\bf Trigger efficiencies:} Trigger efficiencies were measured using
independently triggered events which pass all the analysis cuts by
counting how many did not have a positive response from the primary
trigger.  The $\pipin$ inefficiency was very low ($(12 \pm 4) \times
10^{-4}$) and no correction to $R$ was needed.  The charged trigger
inefficiency was $(832 \pm 9) \times 10^{-4}$ (fig.~\ref{figtrig})
which was partially due to a timing misalignment in a part of the run.
A correction to $R$ of $(+9\pm23)\times 10^{-4}$ is applied for the
difference between the $\pipic$ efficiency for $\ks$ and
$\tau$-weighted $\kl$ events.

{\bf Tagging:} The tagging scheme described above is used to assign
each event to be either $\ks$ or $\kl$ based on whether or not a
proton is seen in the tagger in coincidence with the event.  A
conservative $\pm 2$~ns coincidence window is used which leads to a
probability of erroneously failing to tag a real $\ks$ event ($(1.5
\pm 0.1) \times 10^{-4}$ measured in $\pipic$ events with vertex
tagged events).  Tails in the event time determination which could
lead to a difference in this probability between $\pipic$ and $\pipin$
events and thereby bias $R$ were excluded using events with $\pi^0$
Dalitz decays and photon conversions.  The probability of accidentally
associating a tagger hit with an event is $0.1119 \pm 0.0003)$.  This
is measured separately for $\pipic$ and $\pipin$ events by studying
coincidence windows after artificially displacing the event time.  The
total correction of $R$ due to tagging is $(+18 \pm 11) \times
10^{-4}$.

\section{The result}

\begin{figure}[htb]
  \begin{center}
    \resizebox{0.8\textwidth}{!}{
      \rotatebox{0}{
        \includegraphics[width=\linewidth]{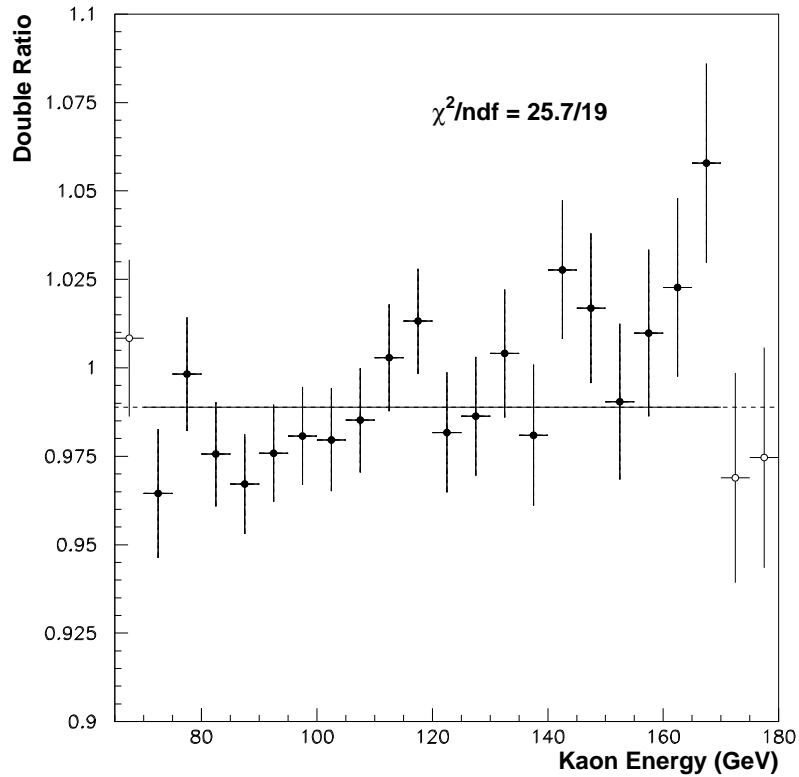}
        }
      }
    \caption{The double ratio in bins of kaon energy.}
    \label{figrvse}
  \end{center}
\end{figure}

\begin{table}[htb]
  \caption{Corrections to {\em R\/}, in $10^{-4}$ units}
  \label{tablecorr}
  \center
  \begin{tabular}{|l|r c r|}
    \hline 
    Tagging                        & $+\,18 $ & $\pm$ & $ 11$ \\
    $\pipic$ trigger efficiency    & $+\,9 $  & $\pm$ & $ 23$ \\
    $\pipin$ background            & $-\;8 $  & $\pm$ & $ 2$  \\
    $\pipic$ background            & $+\,23 $ & $\pm$ & $ 4$  \\
    Acceptance                     & $+\,29 $ & $\pm$ & $ 12$ \\
    Uncorrelated activity          & $-\:2 $  & $\pm$ & $ 14$ \\
    Scattering/regeneration        & $-\,12 $ & $\pm$ & $ 3$  \\
    Neutral scales and linearities & $ \; $   & $\pm$ & $ 12$ \\
    Charged vertex                 & $ \,  $  & $\pm$ & $ 5$  \\
    \hline
    Total correction                & $+\,57 $ & $\pm$ & $ 35$ \\
    \hline 
  \end{tabular}
\end{table}
A summary of the corrections and systematic errors is given in
table~\ref{tablecorr}.  The sample used to compute $R$ contains $4.89
\times 10^5$ $\kl \rightarrow \pipin$ events (accidentally associating
a proton in the tagger with some of these decays means that we only
positively identify $4.34 \times 10^5$ as originating from the $\kl$
beam).  The $\pipic$ trigger was downscaled by a
factor of two, ($\pipic$ events have approximately twice the branching
ratio and twice the acceptance as $\pipin$ events) and the sample
contains $1.071 \times 10^6$ decays.  There are roughly twice the
number of events in each of the $\ks$ decay modes.  The values of the
double ratio $R$ in each bin in kaon energy are shown in
figure~\ref{figrvse}, after
trigger efficiency, tagging, background and acceptance corrections are
included in each bin.  The bins which are used in the analysis,
between 70 and 170~GeV (decided prior to running the experiment), are
shown with black symbols.  The $\chi^2$ of the average is 25.7 (19
d.o.f.).  Extensive checks have been made to exclude systematic biases
which could lead to a variation of $R$ with the kaon energy as the
data might suggest.  As a further check, the double ratio $R$ was
computed in three additional energy bins, as indicated in
figure~\ref{figrvse}.  These three extra points strongly disfavour the
hypothesis of a linear trend in the data.

The result is $R = 0.9889 \pm 0.0027 \mathrm{(stat)} \pm 0.0035
\mathrm{(syst)}$ which, using equation (\ref{eqnr}) leads to
\begin{equation}
\repr = (18.5 \pm 4.5 \pm 5.8) \times 10^{-4}
\end{equation}
The total error is $\pm 7.3 \times 10^{-4}$, obtained by combining the
statistical and systematic errors in quadrature.  This measurement
confirms $\repr$ is non-zero and therefore that direct CP violation
occurs in neutral kaon decays.

\section{Review of direct CP-violation in kaons} 

\begin{figure}[htb]
  \begin{center}
    \resizebox{0.9\textwidth}{!}{
      \rotatebox{0}{
        \includegraphics[width=\linewidth]{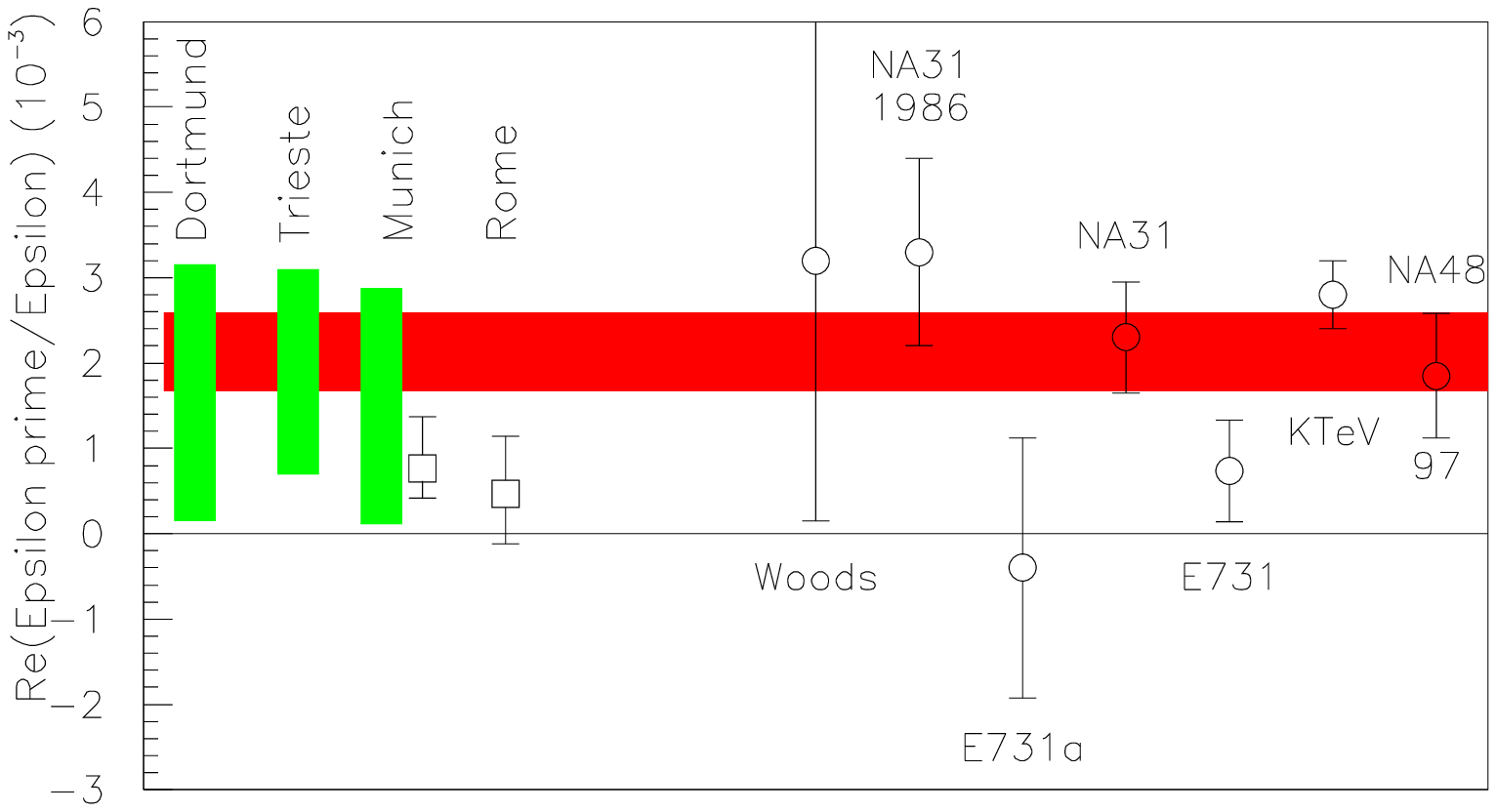}
        }
      }
    \caption{Summary of $\epr$ measurements and predictions.  The predictions
             are made by allowing the parameters to vary over a wide range 
             (shown by the shaded block) or a somewhat narrower range (shown 
             by the error bars on the squares).  The measurements are shown 
             with circles, the NA31-1986 and E731a values are included 
             in the NA31 and E731 values respectively. }

    \label{figeprime}
  \end{center}
\end{figure}

A comparison of the experimental measurements and theoretical
predictions for $\epr$ are shown in figure~\ref{figeprime}. The
technique of all four experiments is to measure the number of decays
of each of the four modes and using cancellations when inserting them
into equation~(\ref{eqnr}) wherever possible to reduce systematics.
The experiments differ in their approaches to how to overcome the
problem that the large difference in lifetime of the $\ks$ and $\kl$
mesons means that they have very different decay distributions in the
experimental apparatus.  The NA48 approach involving taking all four
modes simultaneously and $\tau$-weighting has been described above.
E731\index{E731 experiment} was the first experiment to take data with
all four modes measured simultaneously (about 20\% of their data
sample), the rest was taken with separate runs for the two $\pipic$
modes and for the two $\pipin$ modes.  KTeV\index{KTeV experiment}
collect all four modes simultaneously, although the data presented so
far have also been with separate runs for the two $\pipic$ modes and
for the two $\pipin$ modes.  Both KTeV and E731 use Monte-Carlo
simulation to correct the results for the different acceptances caused
by the different decay distributions of the events.  This technique
maximises the statistics available for the measurement.
NA31\index{NA31 experiment} used a somewhat different technique in
which a $\kl$-like decay distribution of $\ks$ decays was synthesised
by moving the $\ks$ target up and down the fiducial region during the
run.  This required that the two $\kl$ modes were taken at different
times from the two $\ks$ modes. A carefully designed experiment was
used to avoid biases from different detector and trigger rates under
the various beam conditions.

The theoretical predictions from the various
groups~\cite{throme,thmunich,thtrieste,thdortmund} are also shown on
figure~\ref{figeprime}.  The computation is decomposed into
constituent pieces using a product-operator expansion and the
different groups then use a variety of techniques to compute the
individual contributions.  The most difficulty is encountered in
computing a contribution known as $B_6$ which corresponds to the
gluonic penguin diagrams.  Most of the computations find that the
contributions destructively interfere to some extent to yield a
prediction that epsilon prime is smaller than the measured value.
However, one can choose values of the components which are each within
the uncertainty of the computation and which combine to yield a value
consistent with the measurements.  A new computation, completed after
the conference yields a large negative value of $\repr$~\cite{thsoni}.
For a review of the theoretical status of epsilon prime see
also~\cite{thextra}.

\section{Conclusions and outlook}

The first results from the new round of $\repr$ experiments are
confirming that the value is positive.  The average value of $\repr$
is $(21.3 \pm 4.6) \times 10^{-4}$ \index{epsilon prime} when the
errors are scaled in the style of the particle data group~\cite{pdg}.
Both NA48 and KTeV were running at the time of the conference and have
collected large new datasets which are currently being analysed.  The
KLOE\index{KLOE experiment} experiment at DA$\phi$NE is starting and
will be able to provide a measurement~\cite{lpkloe}.  Overall, the
precision to which $\repr$ is known is set to increase
dramatically in the coming years which will open up challenging new
opportunities for theoretical predictions.

\footnotetext[1]{\scriptsize
\textsc{Cagliari:} V.~Fanti, A.~Lai, D.~Marras, L.~Musa;
\textsc{Cambridge:} A.~Bevan, T.~Gershon, B.~Hay, R.~Moore, K.~Moore, 
D.~Munday
% ...
A.~Taurok.
}

%


\def\Discussion{
\setlength{\parskip}{0.3cm}\setlength{\parindent}{0.0cm}
     \bigskip\bigskip      {\Large {\bf Discussion}} \bigskip}
\def\speaker#1{{\bf #1:}\ }

\Discussion

\speaker{Michael Peskin (SLAC)} It is unfortunately easy to overstate
the precision in the theoretical prediction of $\epsilon^\prime$.  One
of the longstanding problems in lattice gauge theory has been in
finding the correct large-distance part of the $\Delta I= \frac{1}{2}$
enhancement in $K^0\to \pi\pi$.  The lattice gauge theory calculation
of $B_6$ has similar difficulties and should be viewed skeptically.
There are new methods in lattice gauge theory that promise to improve
this situation, but results for $B_6$ from these methods are not yet
available.



\begin{thebibliography}{99}

\bibitem{cronin}
J.H.~Christenson {\it et al.},
Phys.\ Rev.\ Lett.\ {\bf 13}, 138 (1964).

\bibitem{cppipin}
M.~Banner {\it et al.},
Phys.\ Rev.\ Lett.\ {\bf 21} 1103 (1968), {\it ibid} 1107. \\
I.~Bugadov {\it et al.},
Phys.\ Lett.\ {\bf 28B}, 215 (1968). 

\bibitem{cpkethree}
S.~Bennett {\it et al.}, 
Phys.\ Rev.\ Lett.\ {\bf 19}, 993 (1967).

\bibitem{cpkmuthree}
D.E.~Dorfan {\it et al.}, 
Phys.\ Rev.\ Lett.\ {\bf 19}, 987 (1967).

\bibitem{cppipicgamma}
E.~Ramberg {\it et al.}, 
Phys.\ Rev.\ Lett.\ {\bf 70}, 2525 (1993)
and 
{\it ibid} 2529.

\bibitem{cppipiee}
A.~Alavi-Harati {\it et al.}
[KTeV Collaboration],
Submitted to Phys.\ Rev.\ Lett.\  \\                       
hep-ex/9908020.

\bibitem{lp99bsession}
Talks by 
K.~Honscheid, 
F.~Takasaki, J.~Dorfan, M.~Medinnis and M.~Paulini in these proceedings.

\bibitem{pleprime}
V.~Fanti {\it et al.}
[NA48 Collaboration],
Phys.\ Lett.\ {\bf B465}, 335 (1999),\\
hep-ex/9909022, CERN/EP 99-114.

\bibitem{hburkhardt}
H.~Burkhardt {\it et al.}
[NA31 Collaboration],
Phys.\ Lett.\ {\bf B206}, 169 (1988).

\bibitem{na31}
G.D.~Barr {\it et al.}
[NA31 Collaboration],
Phys.\ Lett.\ {\bf B317}, 233 (1993).

\bibitem{e731}
L.K.~Gibbons {\it et al.} [E731 collaboration],
Phys.\ Rev.\ Lett.\ {\bf 70}, 1203 (1993).

\bibitem{ktev}
A.~Alavi-Harati {\it et al.}
[KTeV Collaboration],
Phys.\ Rev.\ Lett.\ {\bf 83}, 22 (1999)
hep-ex/9905060.

\bibitem{lp99ktev}
Talk by E.~Blucher in these proceedings.

\bibitem{pdg} 
Particle Data Group, ``Review of Particle Properties'', Eur.\ Phys.\ J. 
{\bf 3} 1 (1998)

\bibitem{throme}
M.~Ciuchini {\it et al.}, 
Z.\ Phys.\ {\bf C68}, 239 (1995)
hep-ph/9501265.

\bibitem{thmunich}
A.J.~Buras, M.~Jamin and M.E.~Lautenbacher,
Phys.\ Lett.\ {\bf B389}, 749 (1996)
hep-ph/9608365.

\bibitem{thtrieste}
S.~Bertolini {\it et al.}, Nucl.\ Phys.\ {\bf B514}, 93 (1998).

\bibitem{thdortmund}
T.~Hambye {\it et al.}, Phys. Rev. {\bf D58}, 14017 (1998).

\bibitem{thsoni}
T.~Blum {\it et al.},
hep-lat/9908025.

\bibitem{thextra}
  Talk by S.~Aoki in these proceedings, Talk by G.~Buchalla at the
  Europhysics Conference, Tampere, Finland 1999, and also the proceedings 
  of the Kaon99 conference in Chicago (1999) (in preparation).
  
\bibitem{lpkloe}
Talk by S.~Bertolucci in these proceedings.

\end{thebibliography}
\end{document}